\documentclass[12pt, preprint]{aastex6}
\usepackage{amssymb,amsmath}
\usepackage{multirow}
\usepackage{colortbl}
\usepackage{longtable}

\begin{document}

\title{Details of Resonant Structures Within a Nice Model Kuiper Belt: Predictions for High-Perihelion TNO Detections}

\author{R. E. Pike\altaffilmark{1}}
\author{S. M. Lawler\altaffilmark{2}}

\altaffiltext{1}{Institute of Astronomy and Astrophysics, Academia Sinica, Taipei, 10617, Taiwan}
\altaffiltext{2}{NRC-Herzberg Astronomy and Astrophysics, National Research Council of Canada, 5071 West Saanich Rd, Victoria, BC V9E 2E7, Canada}

\slugcomment{Accepted to AJ}

\begin{abstract}

We analyze a detailed Nice model simulation of Kuiper Belt emplacement from \citet{brasser2013}, where Neptune undergoes a high eccentricity phase and migrates outward.
In this work, which follows from \citet{pike2017}, we specifically focus on the details of structures within Neptune's mean motion resonances and in the high pericenter population of simulated trans-Neptunian Objects (TNOs). 
We find several characteristics of these populations which should be observable in the distant Solar System in future large-scale TNO surveys as a diagnostic of whether or not this mode of Neptune migration occurred in the early Solar System.  
We find that the leading asymmetric libration islands of the $n$:1 resonances are generally much more populated than the trailing islands.
We also find the non-resonant high-$q$ population of TNOs should have higher inclinations than the low-$q$ population due to the importance of Kozai cycling during their emplacement histories. 
Finally, high-$q$ TNOs should be present in roughly equal numbers on either side of distant mean-motion resonances.
These predictions contrast with predictions from other Kuiper Belt emplacement simulations, and will be testable by upcoming surveys.

\end{abstract}

\section{Introduction}

The Nice Model \citep{NiceModel2005} describes how the four giant planets formed in a metastable state, then chaotically scattered to near their current orbits, dramatically affecting the distribution of the Kuiper Belt in the process.  
In order to robustly test the predictions of this theory describing our Solar System's architecture, both detailed observations and numerical simulations with many thousands of test particles are required.  
The former is provided by the current generation of large-scale surveys such as the recent Outer Solar System Origins Survey \citep[OSSOS;][]{Bannisteretal2016}, which has discovered hundreds of trans-Neptunian Objects (TNOs) and carefully tracked them to precisely measure their orbits.  
For the latter, a variety of numerical simulations have been produced based on the Nice model framework.
These simulations can include a range of dynamical evolutions: large instabilities \citep{nice}; large eccentricities for Neptune \citep{nice,brasser2013}; smooth or grainy migration with moderate Neptune eccentricity \citep{kaibsheppard2016,nesvorny2016}.
The specifics of the migration impart signatures into the dynamical structure of objects in the Kuiper belt.

In \citet[hereafter P17]{pike2017} and this work, we use the output from a large-scale simulation of a Nice Model scenario \citep[hereafter B\&M]{brasser2013}.
Because we know exactly what dynamical processes happened in this model (migration, scattering, resonant sweeping, etc.), the model provides a useful starting point for determining whether or not these processes could have happened in our Solar System's early history. 

In P17, the authors took the output of the B\&M migration model and classified the test particles based on integrating the model end-state for 10~Myr using the dynamical classification scheme of \citet{gladman2008}.
The model particle orbital elements and classifications are publicly available\footnote{DOI : 10.11570/16.0009}.
They then compared the characteristics of different Kuiper belt object populations in the B\&M model to observed populations of TNOs by biasing the simulation using the OSSOS Survey Simulator \citep[as described in][]{Shankman2016}\footnote{Publicly available at http://dx.doi.org/10.5281/zenodo.31297}.
P17 found that the innermost resonances in the simulation are overpopulated relative to observations, while the outermost resonances match observed populations and distributions well, except for the 5:1 resonance which appears from surveys to have a much larger population than simulation predictions.  
The distant scattering and detached populations, when properly biased using a knee- or divot-size distribution, also match survey population and distribution fairly well.
The analysis of P17 suggests that the more distant components of the B\&M Kuiper Belt match reality, so further detailed analysis of these simulated populations may provide valuable predictions that will be testable with upcoming discoveries from large TNO surveys.
This is what we present in this work.  

In the sections that follow, we take the classified output of the B\&M model and further analyze the distant resonant populations and the non-resonant high-pericenter population for predictions which will be observable by large-scale TNO surveys in the future. 
By providing these detailed descriptions of the characteristics of distant TNOs, we hope to facilitate discerning Neptune's migration history from among several theorized migration modes.
In Section~\ref{BMmodel}, we summarize the B\&M migration model \citep[more details are available in P17 and][]{brasser2013}.
In Section~\ref{detailedprop}, we present the details of the distributions of test particles that end up in the $n$:1 and $n$:2 mean-motion resonances with Neptune.
Section~\ref{eandidists} discusses eccentricity and inclination distributions, Section~\ref{libration_section} discusses the libration amplitude distributions and the implications for emplacement, and Section~\ref{twotinos} talks about the populations within the different $n$:1 resonant islands as a diagnostic of Neptune migration.
Section~\ref{kozai_section} focuses on the Kozai resonance within mean-motion resonances, which greatly affects the structure both within (Section~\ref{kozaiinsidemmr}) and outside (Section~\ref{outer}) the strongest mean-motion resonances.
Section~\ref{observables} presents predictions for future TNO surveys that will be able to determine whether or not Neptune migrated outward in a Nice model-style instability similar to the B\&M simulation analyzed here, and compares these predictions with lower-$e$ Neptune migration simulations \citep{nesvorny2016,kaibsheppard2016}

\section{The B\&M Migration Model} \label{BMmodel}

The B\&M simulation is a Nice model type migration with the sun, four giant planets, and a disk of test particles.
This simulation is based on ``Run A'' of the \cite{nice} evolution, which begins after the last giant planet encounter.
Over 200~Myr, Uranus and Neptune smoothly migrate outward and their orbits circularize from an initially high $e$ of 0.3, scattering an eccentric disk of particles initially located with semi-major axes, $a$, between $29 < a < 34$~au.
After the smooth outward migration, the planets and test particles were integrated for an additional 3.8~Gyr under only gravitational forces.
The initial 30,000 test particles were each cloned three times by making small alterations to the mean anomaly of the particles after 1~Gyr and 3.5~Gyr to increase sampling to 270,000 test particles, though many are scattered out of the region of the Kuiper belt.
For more details on the B\&M simulation, see \cite{nice} and \cite{brasser2013}.
As in P17, this work focuses on the scattered portion of the simulation, which includes all particles beyond Neptune and inward of the Oort cloud, $\sim3000$~au.

\begin{figure}
\begin{center}
\includegraphics[width=1.\textwidth]{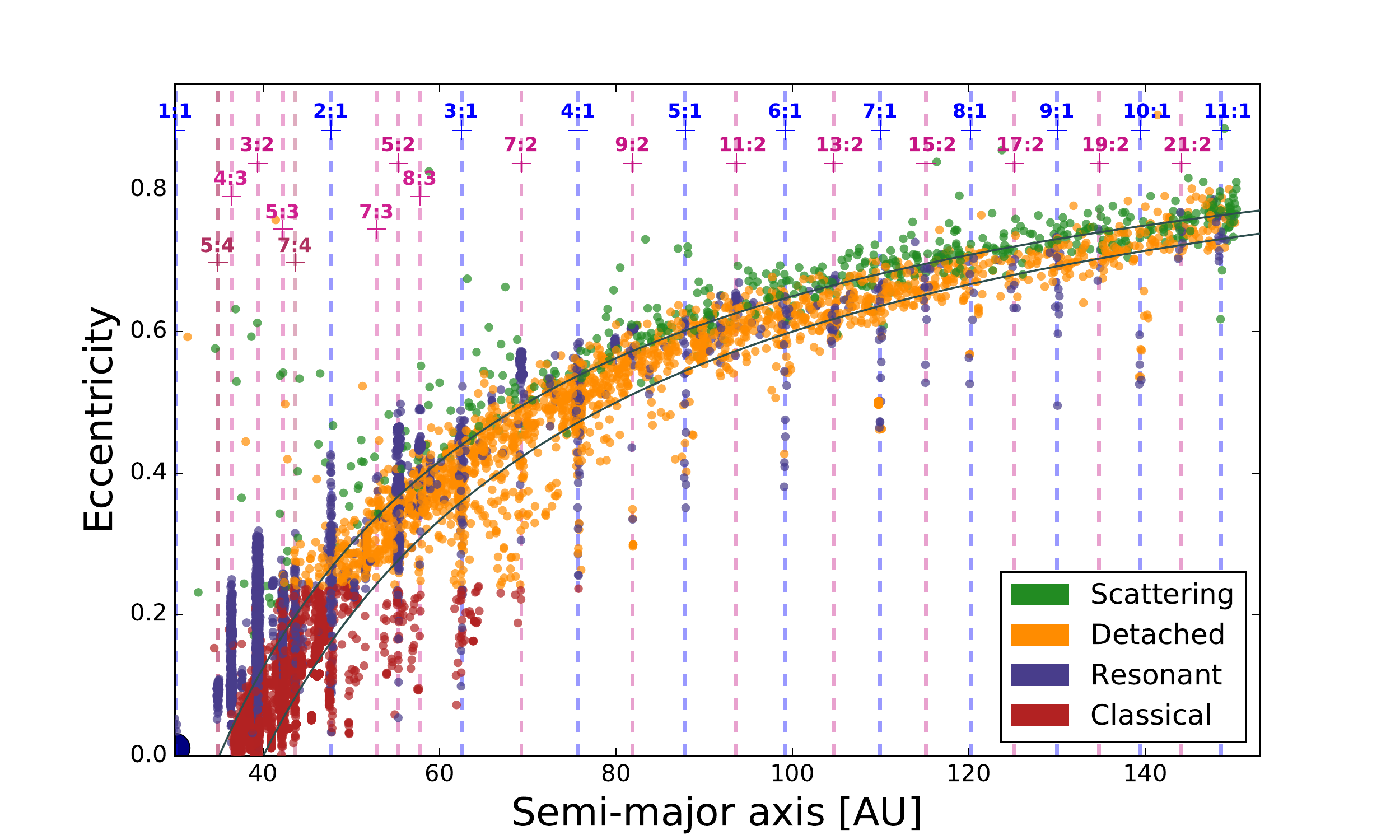}
\caption{
Test particle eccentricity $e$ vs.\ semi-major axis $a$ at the end state of the B\&M model, as classified in P17. The dashed lines show mean motion resonance locations. The solid black lines indicate pericenter distances $q$ of 35 (upper curve) and 40 (lower curve). Neptune is indicated by the large dark blue circle.
}
\label{allparticles}
\end{center}
\end{figure}

The particle classifications, shown in Figure~\ref{allparticles}, are from the P17 analysis of the B\&M model based on integrating the model end state for an additional 30~Myr and characterizing the objects' orbits.
The classification scheme is based on the criteria described by \cite{gladman2008}; for more details on the classification see P17.
For this work, resonant particles were also tested for Kozai oscillation with the spectrogram method \citep{shankmanFFT} on the argument of pericenter, $\omega$, to determine if the test particle $\omega$ oscillates instead of circulating.
A significant fraction of resonant particles were identified as experiencing Kozai oscillations within mean motion resonances.

Because this Nice model simulation includes a high-eccentricity phase for Neptune, Neptune's mean motion resonances (MMRs) are initially wide and powerful, capturing many particles.  
Some particles stay in the resonance as Neptune smoothly migrates outwards and circularizes \citep{nice}, but many fall out of resonance as the resonances become narrower and migrate outwards, and the test particles are forced to higher $e$ and $i$ by Neptune's migration \citep[e.g.][]{Malhotra1993}.
Kozai cycles also play an important role in resonant dropout. 
As particles cycle to lower eccentricities where the MMRs are narrower, dropout becomes more likely \citep{Gomes2011}.
This causes low-$e$ `fingers' near resonances (see discussion in Section~\ref{kozaistructure}).

The relative populations and structure within and near MMRs is a powerful diagnostic of Neptune's migration \citep[e.g.][]{gladman12}.
In this work, we present and analyze the predictions for these structures based on the B\&M Nice model Neptune migration.
Current large scale surveys such as the Outer Solar System Origins Survey \citep[OSSOS;][]{Bannisteretal2016} and future surveys will detect enough TNOs at large distances that detailed predictions from large-scale migration simulations, such as that presented in this work, are necessary to distinguish between different theorized modes of the Solar System's dynamical history.

\section{Detailed Properties of the Resonant Populations} \label{detailedprop}

\subsection{Eccentricity and Inclination Distributions} \label{eandidists}

The eccentricity distribution of the resonances show `fingers' of descending eccentricity from the main population of outer classicals and detached TNOs (Figure~\ref{allparticles}).
These populations are likely created as a result of the Kozai mechanism \citep{Gomes2005}, which also contributes to the hotter inclination distributions.
This effect is stronger for the $n$:1 resonances, but is also apparent for the $n$:2 resonances.
(See Section \ref{kozai_section} for a discussion of the extensive effects of Kozai.)
Simply due to large distances, TNOs in these distant resonances are sparsely sampled by current observations particularly at low-$e$ \citep[e.g.][]{gladman12}, and theoretical models of their distributions would provide useful constraints for population models derived from surveys.

\begin{figure}
\begin{center}
\includegraphics[width=1.\textwidth]{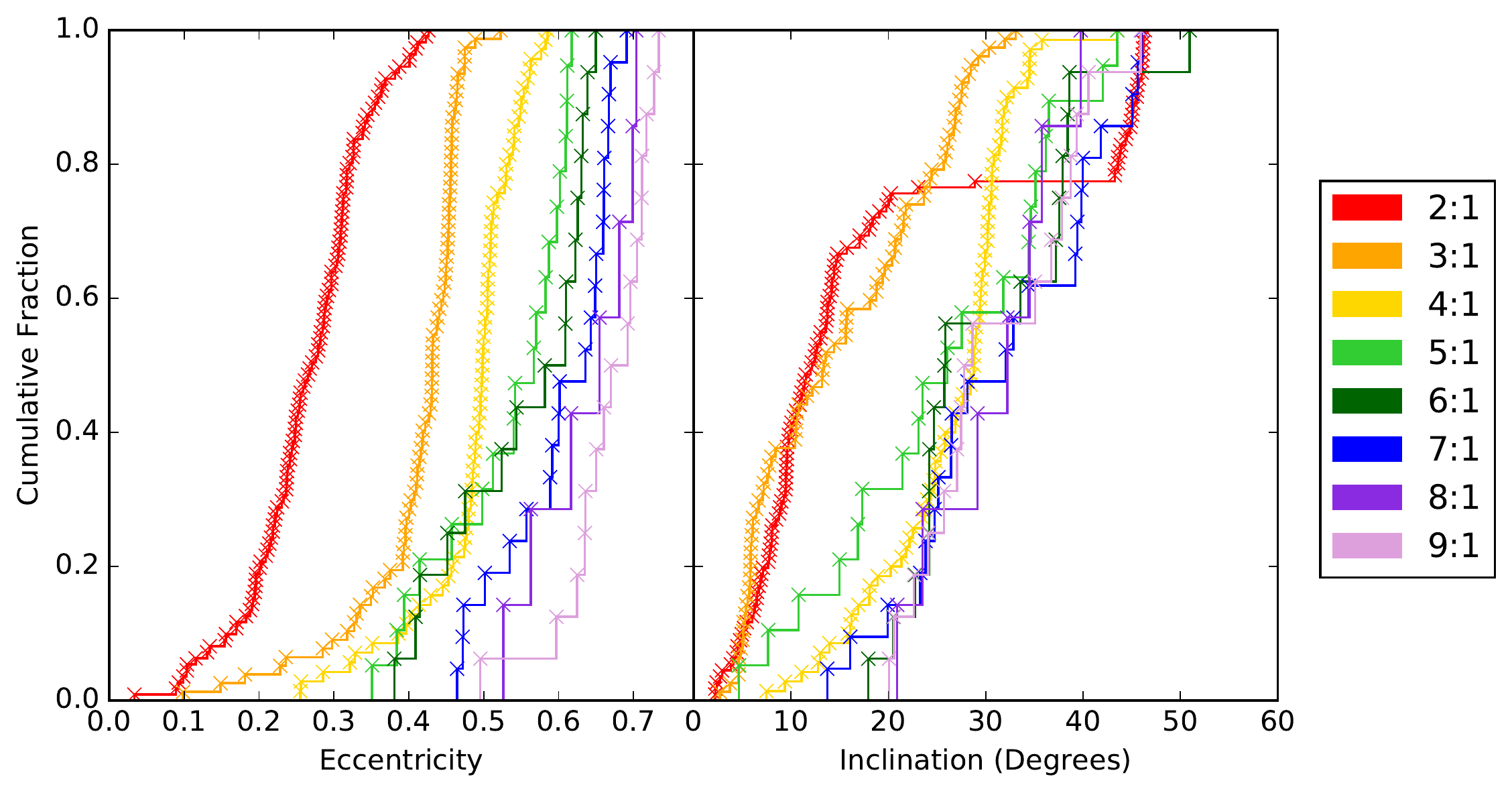}
\caption{The cumulative fractions of the $n$:1 test particles are given for eccentricity $e$ (left) and inclination $i$ (right).  The 2:1 and 3:1 have a significantly different morphology, especially in the $i$-distribution, than the more distant resonances (the high inclination component of the 2:1 is slightly exaggerated due to cloning, see discussion in Section~\ref{twotinos}).  Beginning with the 4:1, there is a slight trend for hotter $i$ values with increasing $a$.  
}
\label{n_one}
\end{center}
\end{figure}

\begin{figure}
\begin{center}
\includegraphics[width=1.\textwidth]{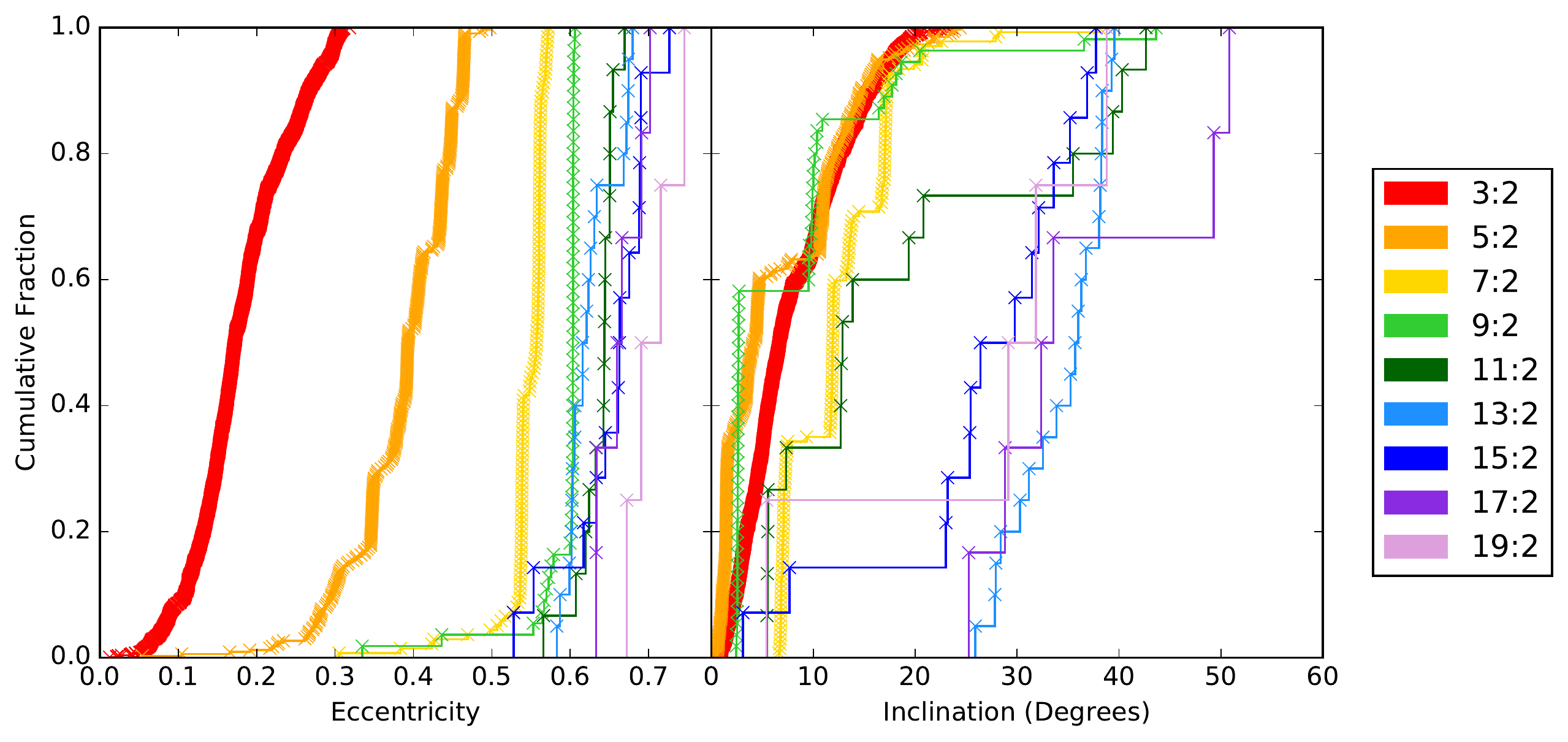}
\caption{As in Figure \ref{n_one}, the cumulative fractions of $n$:2 test particles in the resonances are given for eccentricity $e$ (left) and inclination $i$ (right).  The resonances beyond the 5:2 have $<5$\% of members with $e<0.5$.  The 5:2 and 9:2 contain a significant fraction of low-$i$ members.}
\label{n_two}
\end{center}
\end{figure}

Figures~\ref{n_one} and \ref{n_two} show cumulative distributions in eccentricity and inclination for the $n$:1 and $n$:2 resonances, respectively.  
The resonances exterior to the 2:1 resonance do not contain very low $e$ members, and the minimum $e$ roughly increases with semi-major axis.
The maximum $e$ also increases with semi-major axis; this is an expected result because at larger $a$, a larger $e$ is necessary for the particle to scatter off Neptune.
The inclination distribution of objects in the distant resonances is visibly broader and hotter than the 3:2 and 2:1.
This different morphology between the resonances likely results from the additional Neptune encounters needed to emplace an object at large-$a$.

The range of eccentricities and inclinations displayed by the distant $n$:1 and $n$:2 resonances are particularly interesting, because they affect population estimates and are weakly constrained by surveys \citep[e.g.][]{volk2016}.
Survey population estimates suggest the distant resonances should have relatively large populations, albeit with large uncertainties \citep{gladman12,pike2015,alexandersen}.
Because the highest eccentricity members of a resonance will have the closest pericenter distances and thus brightest apparent fluxes, the highest eccentricity members of any resonance are the objects most likely to be detected in a magnitude-limited survey.
In the more distant resonances, TNOs on low-eccentricity orbits are not ever bright enough to observe in OSSOS and other current surveys, thus the total population of these distant resonances is impossible to constrain unless an eccentricity distribution is assumed.  
The B\&M model presented here suggests a possible eccentricity distribution that is tied to a dynamical history for the Solar System.

Several conclusions are evident from the distributions of inclination and eccentricities for the $n$:1 and $n$:2 populations in the B\&M model.
The B\&M model suggests a hot inclination distribution is likely for all $n$:1 resonances beyond the 3:1, but increased inclinations are not obvious for the $n$:2 resonators until the 13:2, at a much larger semi-major axis.
(The 2:1 resonance has an oddly biased inclination distribution, with an exaggerated high inclination component due to cloning: see further discussion in Section~\ref{twotinos}.)
The average eccentricities of $n$:1 and $n$:2 resonant populations increases with semi-major axis; the more distant $n$:1 resonances all have less than 5\% of particles with $e<0.35$ (see Figure \ref{n_one}) and the distant $n$:2 resonances have less than 5\% of particles with $e<0.5$ (see Figure \ref{n_two}).
The bulk of the population in these resonances has nearly the maximum stable eccentricity.
If this distribution represents reality, population estimates which exclude a low-$e$ component are not significantly underestimating the population \citep[e.g.][]{volk2016}, while a uniform $e$-distribution would overestimate the population by including far too many TNOs at low-$e$.

The $n$:1 resonances tend to have higher inclinations and eccentricities at a given semi-major axis than nearby $n$:2 resonances.
We suspect this may be because capture into the $n$:1 resonances requires more scattering events than the $n$:2's, although previous simulations \citep{LykawkaMukai2007Scattering} find that capture probability is highest for the lowest-order resonances, and also for resonances at larger $a$.
Whether this effect is due to capture probabilities or to the timescale of diffusion to higher $e$ and $i$ within the resonances, and whether or not Kozai oscillations play a significant role, will require further modeling.

\subsection{Resonance Libration}
\label{libration_section}

\begin{figure}
\begin{center}
\includegraphics[width=1.\textwidth]{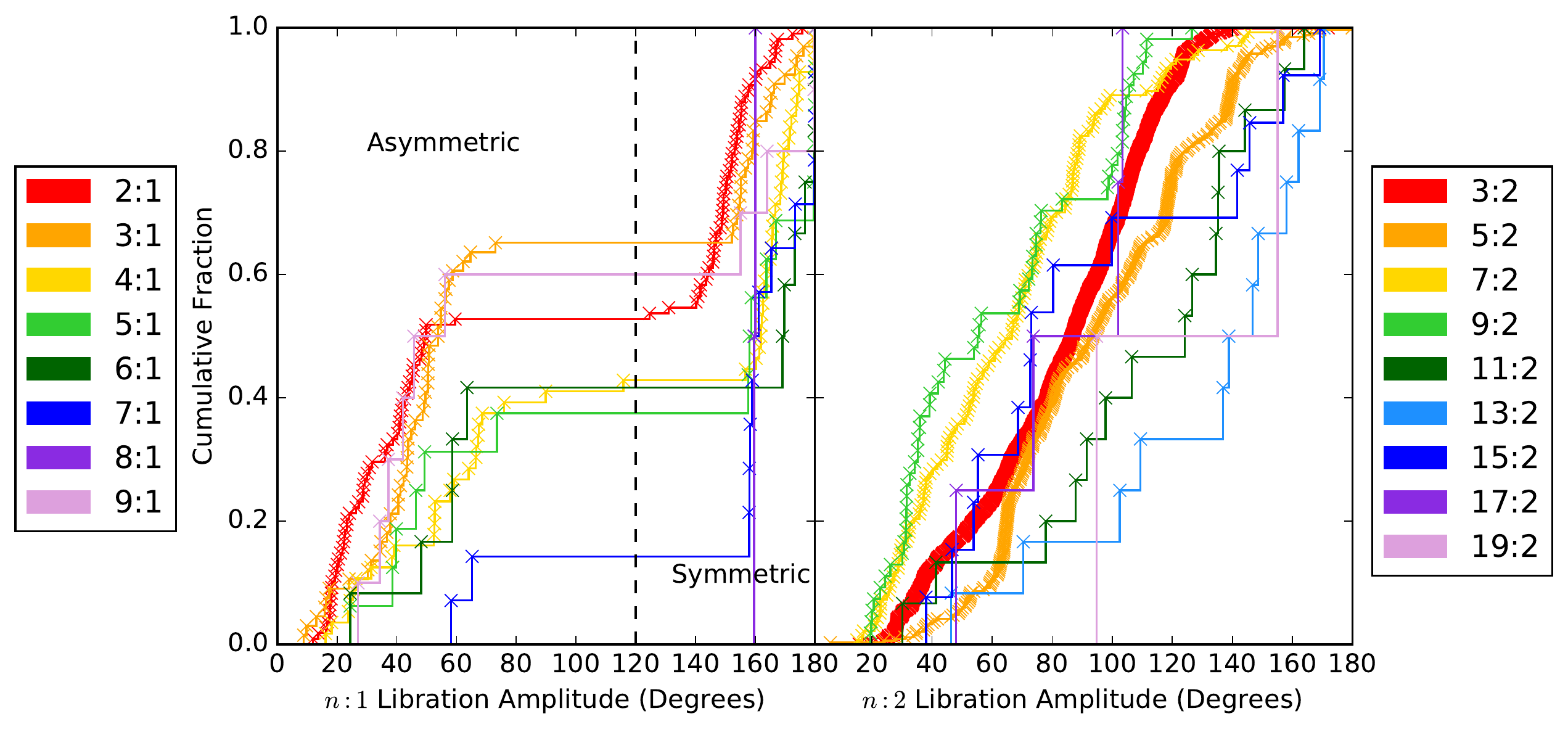}
\caption{The cumulative fractions of libration amplitudes for the stable test particles in the $n$:1 (left) and $n$:2 (right) resonances.  The $n$:1 resonators include high amplitude symmetric librators and a high fraction of low amplitude asymmetric librators.  The $n$:2 includes only symmetric resonators, which possess a large range of libration amplitudes.
}
\label{n12_libration}
\end{center}
\end{figure}

The libration amplitude, or range of resonant angles a particle explores (the amplitude of Equation~1 in P17), is influenced by the specific capture mechanisms and migration timescales \citep[e.g.][]{chiang2002}.
The libration amplitudes of the stable B\&M $n$:1 and $n$:2 resonators are shown in Figure \ref{n12_libration}.
The $n$:2 resonators occupy a single resonant island around 180$^{\circ}$ and have libration amplitudes from 6$^{\circ}$ (for the lowest libration amplitude 5:2 resonator) to just less than 180$^{\circ}$ (for the largest amplitude 5:2 resonator).
The $n$:1 test particles include both large-amplitude symmetric and small-amplitude asymmetric librators.

The B\&M $n$:2 resonators have a large range of libration amplitudes in a single libration island.
The 3:2 and 5:2 resonances have roughly uniform distributions from 20$^{\circ}$ to 140$^{\circ}$ and 160$^{\circ}$.
The 7:2 and 9:2 resonances have a lower median libration amplitude, and the 11:2 and 13:2 resonators have a higher mean libration amplitude than the 3:2 and 5:2 resonators.
In all of these resonances, libration amplitudes below 20$^{\circ}$ are either nonexistent or an insignificant fraction of the population.
This reflects a combination of reduced phase space at low amplitude and the inefficiency of deep capture because these populations are implanted through scattering capture and not through sweep-up.
\cite{volk2016} searched for low-libration amplitude Plutinos (3:2 resonators) in the OSSOS survey. 
Their minimum libration amplitude detected TNO has an amplitude of 10$^{\circ}$$^{+8}_{-4}$, which may require resonance sweeping capture.
This low libration amplitude Plutino discovery suggests that the Plutinos include a sweep-up capture component; the presence or absence of low-libration amplitude 5:2 resonators in the real Kuiper belt could provide useful insight into the extent of the primordial planetesimal disk.
If a significant low-libration amplitude component exists in the 5:2, the primordial disk would have had to extend to large enough distances that sweep-up could operate effectively in the 5:2, suggesting a much wider primordial disk than often theorized. 
A small, swept-up, low libration amplitude component resulting from a lower density extended disk (not included in B\&M) would be consistent with the Neptune's migration evolution in this simulation, but quantifying this component would require additional simulations of the instability.

\subsection{$n$:1 Resonant Island Occupation: Leading Island Enhancement}
\label{twotinos}

The dynamical behavior of the $n$:1 resonators is more complex than the $n$:2 test particles because the $n$:1 resonators can occupy three libration islands: the symmetric island centered at 180$^{\circ}$ as well as the leading (centered at small angles $<$180$^{\circ}$) and trailing (centered at large angles $>$180$^{\circ}$) asymmetric libration islands.
Because of the segregated phase space, symmetric librators all have large libration amplitudes ($\gtrsim120^{\circ}$) and asymmetric librators have smaller amplitudes.
As a result, the relative fractions of symmetric and asymmetric resonators are apparent in Figure \ref{n12_libration}, as a large gap separates the libration amplitude of the symmetric and asymmetric islands.

The fraction of TNOs in the different islands, particularly in the 2:1, has been theorized to be a diagnostic of the timescale of Neptune's migration \citep[Section~\ref{resfractions};][]{chiang2002}.
Due to the large number of test particles in our simulation, we are able to measure this fraction for 6 of the $n$:1 resonances (Table~\ref{n_outer}).
In all of these resonances, the leading asymmetric islands contain more particles than the trailing.
To assess the significance of these values, we compute the probability of randomly finding \emph{at least} the measured number of leading librators when an equal number of particles are expected in the leading and trailing islands (`Likelihood' column in Table~\ref{n_outer}).
Using this, we show that most of the $n$:1 resonances have leading island occupation that is much too high to be explained by random fluctuations.  
The leading islands are more populated in each of the $n$:1 resonances due to emplacement during this simulation.

\begin{deluxetable}{c c c c c}
\tablecaption{Libration Islands of B\&M $n$:1 Resonators
\label{n_outer}}
\tablehead{
$p$:$q$ & Total & Leading & Trailing & Likelihood \\
 & particles & particles & particles &  
}
\startdata
2:1	&	108	&	47	&	21	&	0.1\%	\\
3:1	&	66	&	35	&	13	&	0.1\%	\\
4:1	&	56	&	21	&	12	&	8\%	\\
5:1	&	16	&	6	&	5	&	50\%	\\
6:1	&	12	&	6	&	2	&	14\%	\\
7:1	&	14	&	7	&	0	&	0.8\%	\\
\enddata
\tablecomments{``Likelihood'' column gives the probability of randomly drawing the measured number of particles or more in the leading island when the expectation is for an equal number of particles in the leading and trailing libration islands.}
\end{deluxetable}

\subsubsection{Leading Island Enhancement is not an Artifact of Integration Methods}

To verify that this inflated leading island occupation is not an artifact of integration time-step selection, we perform a series of long-timescale integrations of 2:1 resonators across a wide swath of the resonant phase space with different time-steps, ranging from 0.5 to 0.04 years (Table~\ref{21timesteps}).
We used 10,000 randomly generated test particles over a range of semi-major axes, inclinations, and eccentricities which cover the 2:1 resonant phase space; approximately 900 of these test particles were resonant at the start of the integrations.
We find that the fraction of particles in the leading and trailing islands is approximately equal at the beginning of the simulation, and that the fluctuations in island occupation are consistent with random, as evidenced by the high likelihoods calculated in Table~\ref{21timesteps}.
These likelihoods are calculated in the same manner as in Table~\ref{n_outer}, that is, the probability of randomly drawing the at least the observed number of particles observed in the leading island when the expectation is for 50\% of the asymmetric librators to be in the leading island.

\begin{deluxetable}{c c c c c c}
\tablecaption{Libration Islands of Randomly Populated 2:1 Resonators with Differing Timesteps \label{21timesteps}}
\tablehead{
Time-step	&	Time-step	&	Total	&	Leading	&	Trailing	&	Likelihood \\
size	&	&	particles	&	particles	&	particles	&	 \\
}
\startdata
0.5 yr	&	0 Gyr	&	867	&	211	&	220	&	68\%	\\
	&	1 Gyr	&	384	&	85	&	101	&	89\%	\\
	&	2 Gyr	&	317	&	75	&	91	&	91\%	\\
	&	3 Gyr	&	178	&	53	&	63	&	85\%	\\
	&	4 Gyr	&	230	&	53	&	69	&	94\%	\\ \hline
0.1 yr	&	0 Gyr	&	871	&	217	&	222	&	61\%	\\
	&	1 Gyr	&	373	&	81	&	104	&	96\%	\\
	&	2 Gyr	&	296	&	74	&	87	&	87\%	\\
	&	3 Gyr	&	186	&	59	&	58	&	50\%	\\
	&	4 Gyr	&	214	&	52	&	62	&	85\%	\\ \hline
0.04 yr	&	0 Gyr	&	877	&	217	&	224	&	65\%	\\
	&	1 Gyr	&	371	&	84	&	102	&	92\%	\\
\enddata
\tablecomments{``Likelihood'' column is computed as in Table~\ref{n_outer}.}
\end{deluxetable}

Examining the fractional changes in resonant island occupation (Table~\ref{21timesteps}) allows us to estimate how this over-representation compares to jitter within the occupation rates of the islands due to normal perturbations.  
From inspection of Table~\ref{21timesteps}, it appears that over time, simply by random diffusion and stability, the occupation of the trailing island is slightly higher than the leading island, and this occurs regardless of the integration time-step (however this slight trailing overpopulation is not statistically significant according to our likelihood calculations, and is consistent with 50\% occupation).
This weak trend toward trailing island overpopulation is the opposite of the statistically robust leading island overpopulation that we find in the B\&M model.
Long term evolution, even with different integration time-steps, cannot explain the overpopulation observed in the $n$:1 resonances of the B\&M model.

\subsubsection{Leading Island Enhancement is not Caused by Neptune's Orbital Circularization}

\begin{figure}
\begin{center}
\includegraphics[width=1.\textwidth]{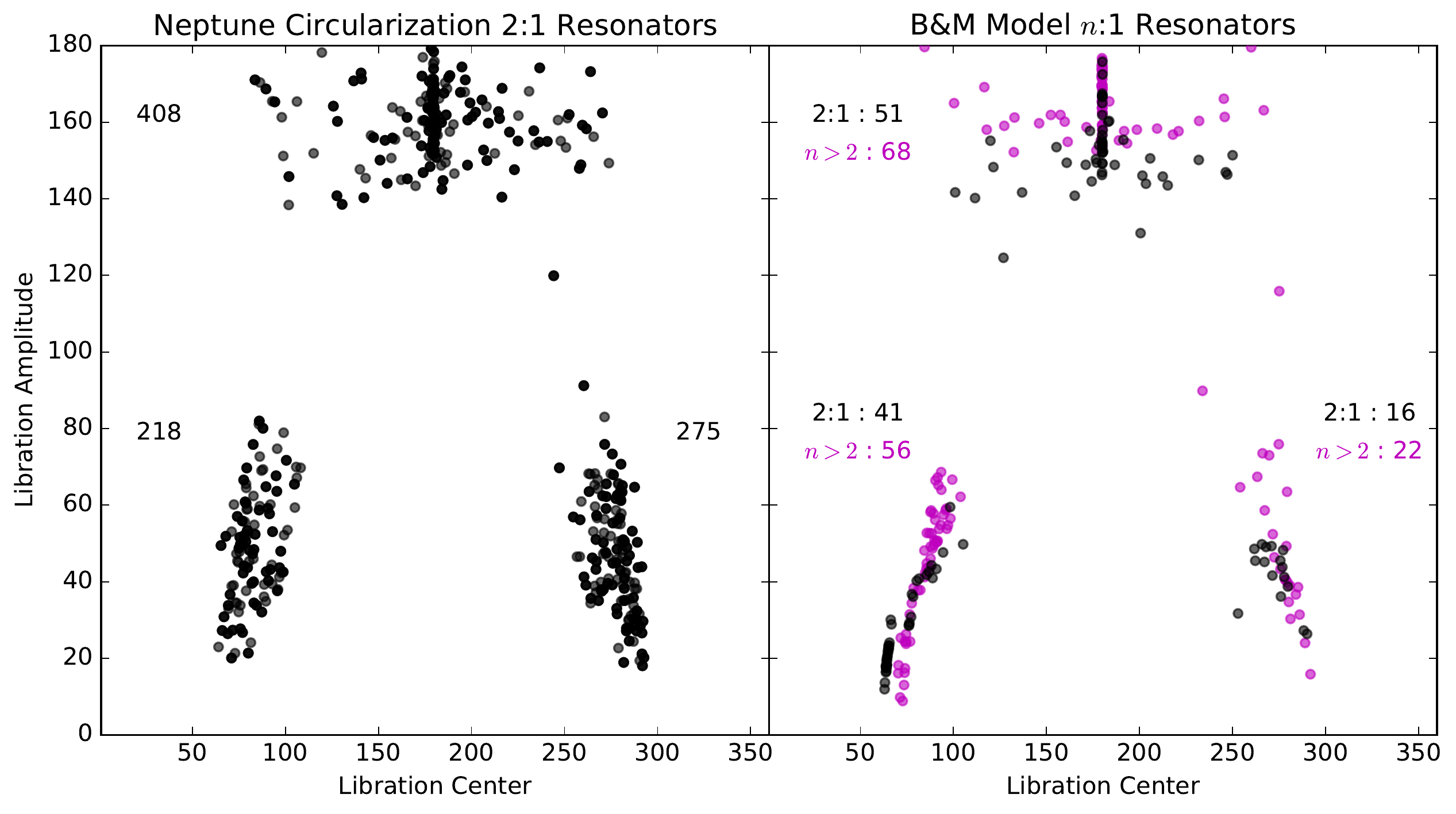}
\caption{Left: The 2:1 resonators from the simulation of the circularization of Neptune's orbit.  The numbers indicate the number of particles in each island.  There is a 26\% enhancement of the trailing libration island compared to the leading libration island, which has a statistical likelihood of only 0.5\% if equal occupation is expected.  This is the \emph{opposite} of the enhancement seen in the B\&M simulation.  Right: The B\&M simulation $n$:1 resonators.  Stable 2:1 resonators are shown in black, and stable 3:1, 4:1, 5:1, 6:1, and 7:1 resonators are shown in magenta.  A simple circularization of a large eccentricity Neptune does not produce this signature.}
\label{asymmetric}
\end{center}
\end{figure}

We explored the effect of the circularization of Neptune's orbit on the populations in the leading and trailing 2:1 resonant islands with an additional simulation.
This simulation used an initial Neptune eccentricity of 0.3, damped to $e \sim 0.01$ over 10~Myrs.
The other giant planets were included with their current orbital configuration.
26,450 test particles were given initial conditions covering the phase space of the 2:1 resonance.
After Neptune's orbit circularized, the end state of the simulation was integrated without migration for an additional 10~Myrs for classification, as we did for the B\&M simulation.
901 particles were identified librating in the 2:1 resonance, with 218 in the leading and 275 in the trailing libration island, a 26\% enhancement of the trailing island (Figure \ref{asymmetric}), the \emph{opposite} enhancement from that seen in the B\&M simulation.
If equal populations are expected in both islands, this enhancement has a statistical likelihood of only 0.5\%.
This migration simulation involving only Neptune circularization results in a statistically significant enhancement of the trailing libration island.

The test particle initial conditions were selected to initially populate the 2:1 resonance and lose resonators only as a result of Neptune's circularization, with no outward migration. 
This evolution did not result in the enhancement of the leading island seen in the B\&M simulation; as shown in Figure \ref{asymmetric}, it resulted in the opposite enhancement.
The full emplacement scenario is clearly important for creating the B\&M asymmetry; it cannot be attributed to any isolated portion of the migration and emplacement simulation.
The particle initial conditions may also play a role in their resonant behavior, or perhaps the scattering process in B\&M combined with the large $e$ of Neptune and the particles is necessary to enhance the leading island.
The circularization of the large-$e$ Neptune clearly can play a role in creating asymmetries in resonances, but that alone does not reproduce the enhancement in the leading island of the 2:1 resonance.
Further simulations must be undertaken to understand the enhancement effect we observe here.  

\subsubsection{Leading Island Enhancement is not an Artifact of Cloning}

We also carefully checked to make sure that cloning in the B\&M simulation does not significantly affect our results.
Looking at the test particles that end up in the 2:1 resonance, we find that those that are clones of the same original particle are not any more likely to be in the same resonant island than any other particle in the resonance, with one exception.
Several original particles were deeply trapped in a very stable portion of the leading asymmetric island of the 2:1 at rather high inclination and eccentricity.
When these particles were cloned, many of the clones appear to have also remained in this very stable pocket.  
Because three original (uncloned) particles were emplaced in this stable pocket independently, we believe this over-representation is a real effect of this mode of Neptune dynamical evolution, though slightly exaggerated in the final particles due to cloning.
This clump of stability at high $e$ and $i$ and low libration amplitudes, though oversampled by chance in the B\&M simulation, is also found by a detailed stability analysis of the 2:1 resonance \citep{TiscarenoMalhotra2009}, and could be a real structure worth looking for in large TNO surveys.
Cloning the test particles which are stably captured into resonance instead of evolving chaotically can falsely enhance a stable population, however the consistent enhancement in the leading libration island for multiple $n$:1 resonances is a strong indication that this enhancement is a real signature of this migration and not an artifact of cloning.

This leading island overpopulation must be a result of the specifics of Neptune's migration and the test particle initial conditions.
It is not a result of random fluctuations over 4~Gyr of integration, or integration technique.
However, the specific dynamical processes which creates this enhancement are still not clearly identified.

\clearpage{}

\section{The Importance of the Kozai Mechanism}
\label{kozai_section}

The Kozai mechanism results in an exchange between eccentricity and inclination, conserving the $z$-component of angular momentum, $L_z$ \citep{kozai1,kozai2}.
Outside of mean-motion resonances, this can only operate for very high inclinations \citep{thomas1996}.
Inside mean-motion resonances, even moderate to low inclination objects can experience this effect \citep[e.g. Pluto at $i\simeq17^{\circ}$;][]{williams1971}, but at lower amplitudes \citep{morbidelli2005}.
The angular momentum exchange results in anti-correlated oscillations of eccentricity and inclination, and oscillation of the argument of pericenter, $\omega$, all having the same period ($\sim$few Myr timescales).

\subsection{Kozai Influences Structure within Resonances} \label{kozaiinsidemmr}

\begin{figure}
\begin{center}
\includegraphics[width=1.\textwidth]{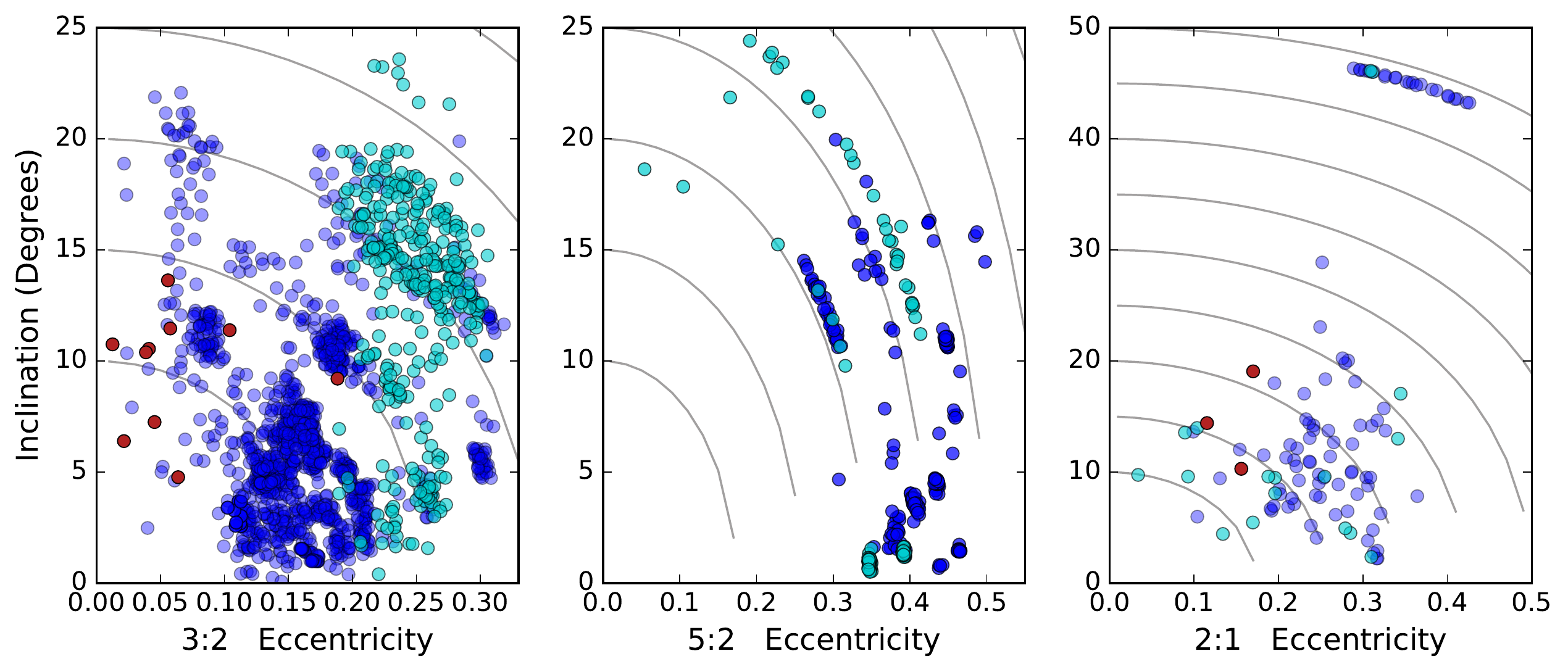}
\caption{The B\&M simulation particles in the 3:2, 5:2, and 2:1 resonances are shown in $e$--$i$ space.  The 3:2 and 5:2 have strong $e$--$i$ dependence.  The structure in the 5:2 and 2:1 resonances (in particular the clump in the at high $e$ and $i$ in the 2:1) is exaggerated as a result of cloning; the 3:2 Kozai particles show the expected continuum distribution.  Grey lines show curves of constant $L_z$ (eq.~\ref{kozaiLz}), which particles move along during a Kozai cycle.  The red circles are not stable in the resonance, the blue circles are stable, and the cyan circles are stable Kozai resonators.  At the end of the B\&M simulation 20--30\% of objects in the 3:2, 5:2 and 2:1 resonances exhibit Kozai oscillations.}
\label{kozai}
\end{center}
\end{figure}

Kozai resonators are identified within several mean motion resonances in the B\&M simulation.
Figure \ref{kozai} shows the eccentricity-inclination distribution for three mean-motion resonances with Kozai oscillators: the 3:2, 5:2, and 2:1 resonances.
Kozai oscillators in each of the resonances occupy a significant fraction of the highest $e$ and $i$ space.

Over the course of a Kozai cycle, as $e$ and $i$ cycle to higher and lower values, the quantity 
\begin{equation} \label{kozaiLz}
L_z \propto \cos{i} \sqrt{1-e^2}
\end{equation}
is conserved.  
Lines of constant $L_z$ are shown in Figure~\ref{kozai}.  
Some particles in the simulation that were cloned while in a Kozai cycle are distributed along curves of constant $L_z$. 
This is particularly obvious in the 5:2 and 2:1 distributions (see cloning discussion in Section~\ref{twotinos}). 

Particularly for the Kozai Plutinos (3:2 resonators), larger $i$ correlates with a constrained range of $e$ values (see Figure \ref{kozai}).
This dependence is not usually included in population models, which is a major weakness of typical resonant population modeling. 
21\% of the 1640 B\&M Plutinos are in Kozai which agrees well with the highest likelihood Kozai fraction of 20\% for the OSSOS survey found by \cite{volk2016}.
Though the Kozai fraction matches observations, the inclination distribution within the B\&M Plutinos doesn't include sufficiently excited particles to match known Plutinos.
A slower migration rate for which Kozai is more effective may be key to a sufficiently hot inclination distribution for the Plutinos \citep[as in][]{kaibsheppard2016}.

The inclination distribution of the B\&M model 5:2 resonators appears to have multiple discrete components as a result of Kozai resonance and particle cloning during the simulation.
The $e<0.3$ particles all have large $i$; two components for both the eccentricity and inclination are apparent.
29\% of the 337 5:2 resonators are in Kozai.
The inclination distribution of the non-Kozai fraction of the 5:2 is less dynamically hot than the 5:2 population as a whole; this distinction has important implications for population models.

Even though Kozai is not operating for all of the resonant particles at the end-state of the simulation, its effect is clearly seen in the distribution.
The interdependence of inclination and eccentricity as well as the presence of high inclination particles is indicative of Kozai evolution.
As discussed below, Kozai resonance dropout may be just as important for shaping the orbital element distribution within resonances as resonant dropout is for shaping the distribution of detached objects \citep{nesvorny2016,kaibsheppard2016}.

\subsection{Signatures of Kozai Beyond the Main Classical Region}
\label{outer}

\begin{figure}
\includegraphics[width=.95\textwidth]{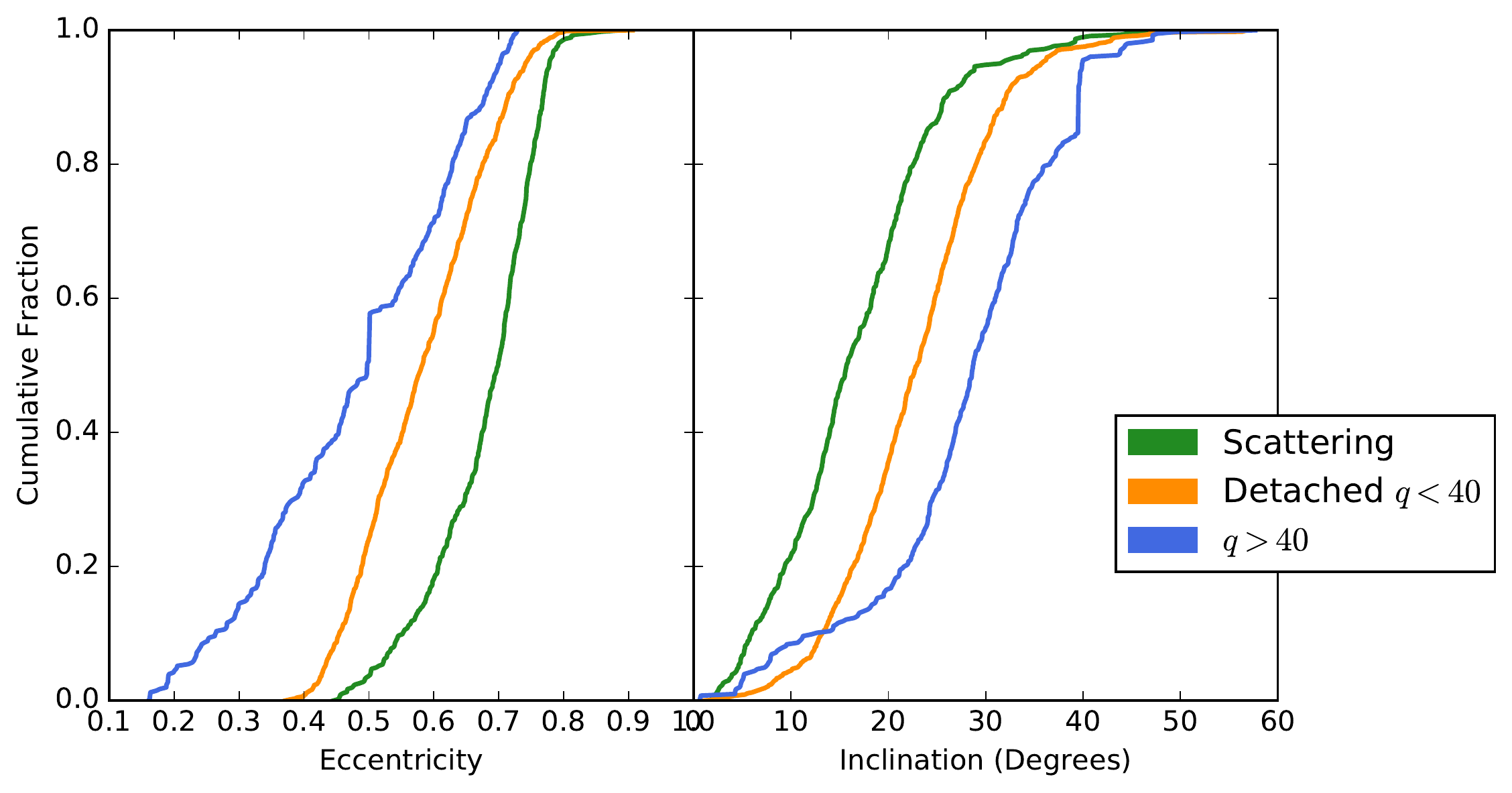}
\includegraphics[width=.83\textwidth]{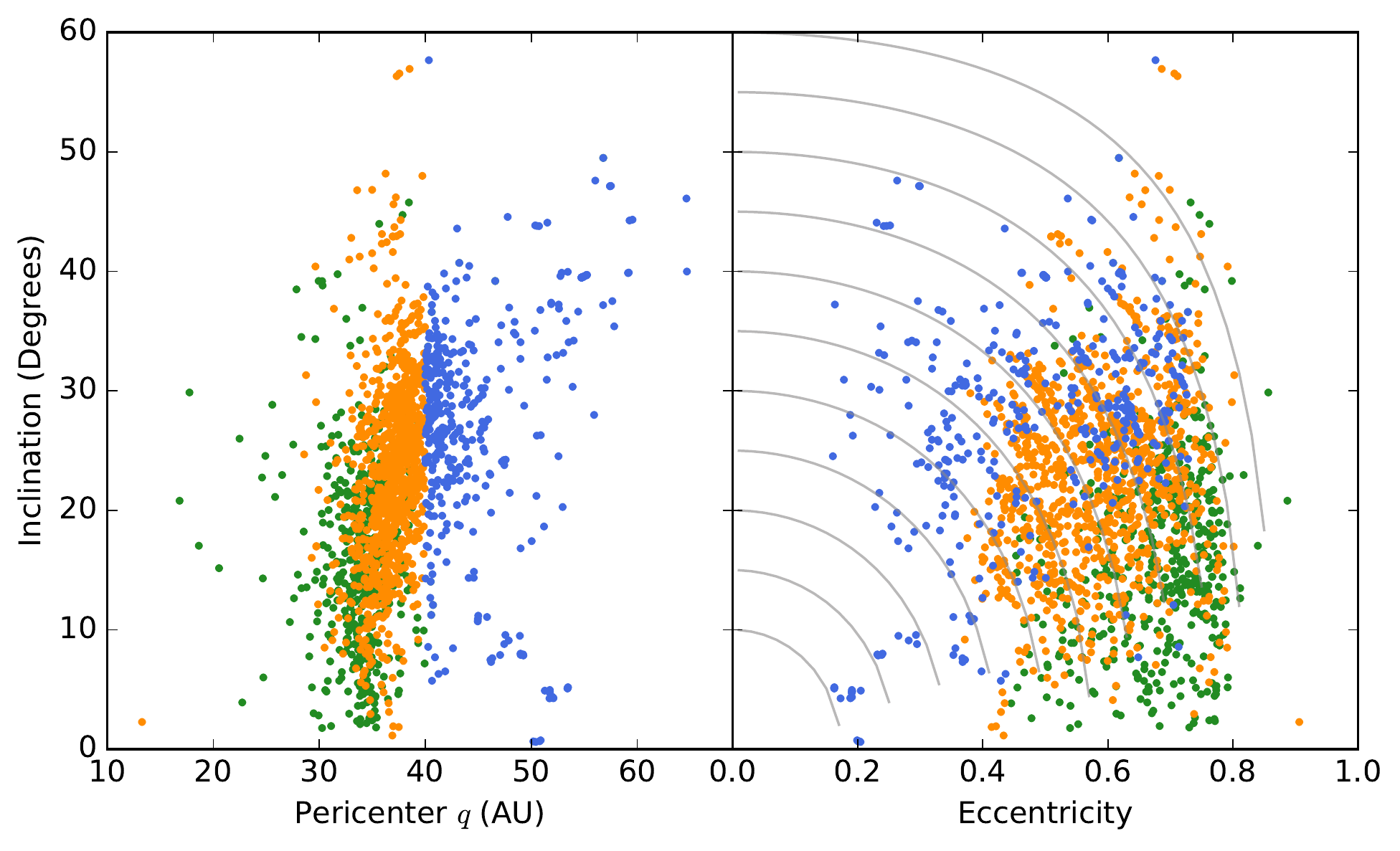}
\caption{Upper: Two panels show the cumulative fractions of non-resonant B\&M objects beyond the 3:1 resonance (62.5 AU) in eccentricity and inclination.  The scattering objects (green) have hotter eccentricities and colder inclinations, while the detached objects with $q<40$~AU (orange) have intermediate inclinations and colder eccentricities.  The high-$q$ particles (blue) have lower eccentricities and higher inclinations. Lower: The anti-correlation of $i$ and $e$ is visible in the $i-q$ plot (left).  The gray lines on the $e-i$ plot (right) are the lines of constant $L_z$ (Equation~\ref{kozaiLz}), as in Figure \ref{kozai}, this structure is evident in the high-$q$ objects and to a lesser extent in the detached $q<40$ objects.
}
\label{distant}
\end{figure}

The effects of temporary evolution in Kozai resonance are apparent in the regions beyond the main classical belt.
In the B\&M simulation, all particles in this region have been emplaced as a result of the planetary migration; the initial planetesimal disk was truncated at 34~AU.
The classification of these particles is based on \cite{gladman2008}.
As shown in Figure~\ref{allparticles}, this region is populated by outer classical objects (at $e<0.24$, assumed to be an extension of the main classical belt in this classification scheme), objects in resonance with Neptune, detached objects, and scattering objects, which have semi-major axes that are changing on short timescales as a result of Neptune's influence.
In Figure \ref{allparticles}, the outer classical objects and some detached objects can be seen located near the resonance fingers.
The $q=40$~AU boundary is shown as a black arc starting at $a=40$~AU.
These outer classical particles all have $q>40$~AU and were likely emplaced by the same mechanism that created the detached objects in the same $q$ range; these particles cannot be primordial.
The high-$a$ ($a>62.5$, outside the 3:1 MMR) particles with $q>40$~AU, which we refer to as `high-$q$' objects, seem to be a more informative dynamical subdivision and include test particles from the detached and outer classical populations.
These subpopulations have different dynamical characteristics as a result of their emplacement histories.
Figure \ref{distant} shows the inclination and eccentricity distributions of particles in the various sub-populations in the distant Kuiper belt.

The \cite{gladman2008} dynamical classification system beyond the main classical belt is not optimal for understanding the B\&M simulation results.
This classification separates test particles which share a common origin into different sub-components, and it also places into the same classification some particles with very different evolutionary histories.
The `outer classical' objects in the B\&M simulation are clearly just the low-$e$ extension of the high-$q$ population, which is much more dynamically distinct.
With the initial conditions of the B\&M model, objects in the outer classical region cannot be primordial.
There is a shared dynamical origin for the outer classical and some detached objects, so a pericenter cut for dynamical classification is more informative than the \cite{gladman2008} classification in the distant Solar System.

Figure~\ref{distant} shows that the high-$q$ subpopulation has different characteristics than the detached and scattering populations.
Because of their selection criteria, these high-$q$ particles have a much colder eccentricity distribution, reaching down to $e=0.16$.
However, this population also has a hotter inclination distribution, with a median of 28$^{\circ}$.
Based on the anti-correlated $e$ and $i$ distributions, it appears that these particles were emplaced through resonance dropout or diffusion, after the Kozai mechanism reduced their eccentricities and increased their inclinations.
In the $e$-$i$ plot in Figure \ref{distant}, the lines of constant $L_z$ resulting from Kozai resonance are apparent.
When objects evolve to low-$e$, where the resonance is narrower, particles are more likely to exit the resonance.
Some of the detached objects are not particularly close to a resonance, and these particles (as well as others near resonance) may have been emplaced through dropout as Neptune (and the resonances) migrated outward \citep{gomes2003}.
A circularization of Neptune's orbit can also shrink the width of the resonance, particularly at low eccentricity, and cause resonance dropout \citep{gomes_ssbn}.
If particles drop out during migration, they should be primarily located Sunward of resonances, while dropout during circularization would have a more symmetric signature around resonances (see Section~\ref{kozaistructure}).
These high-$q$ particles are consistent with formation through resonant dropout after modification by Kozai within distant mean-motion resonances.

The detached population includes the majority of these high-$q$ objects in addition to a large $q<40$~AU population.
The eccentricity distribution of the detached objects is somewhat hotter and the inclination distribution is somewhat colder than the $q>40$~AU sub-population, indicating the larger sample includes particles less modified as a result of Kozai.
In Figure \ref{distant}, only the $q<40$~AU sub-component of the detached objects is presented.
These particles still may be emplaced through resonant dropout, however the Kozai mechanism likely did not play a significant role in the $q<40$~AU portion of the detached population.

The scattering objects, in contrast, have hotter eccentricity distributions and colder inclination distributions.
These particles were also likely modified by the Kozai mechanism, increasing the eccentricity and decreasing the inclination during a resonance capture.
The mean-motion resonances are weaker at the extreme values of $e$ and $i$, so escape is more likely for these particles.
The lines of constant $L_z$ are not apparent in this population in Figure \ref{distant}, because the scattering process erases this signature.
After the particle diffuses or drops out of resonance it is scattered by Neptune on relatively short timescales.

The distant particles in the B\&M simulation have a range of eccentricity and inclination distributions indicative of their emplacement history.
The Kozai mechanism within mean motion resonances has resulted in populations anti-correlated in $i$ and $e$, which is evident in the plot of all objects in $i$ vs. $q$ in Figure \ref{distant}.
This distribution of objects in the B\&M simulation in $e$ and $i$ is consistent with the observed Kuiper belt, and the effect of the Kozai mechanism has been noted in previous simulations, which found a noticeable fraction of distant resonant objects evolving to low-$e$ \citep{gomes2003,pike2015,nesvorny2016,kaibsheppard2016}.

\section{Observables which test Neptune's Mode of Migration} \label{observables}

\subsection{Resonant Island Fractions}
\label{resfractions}

The relative fraction of leading and trailing asymmetric librators for the 2:1 resonance may be diagnostic of Neptune's mode of migration.
In a different Neptune migration scenario with a low eccentricity Neptune, \cite{chiang2002} tested a range of simulation migration speeds, from 50,000 years to 5~Myr, and found that the faster migration speeds increased the fraction of 2:1 resonators in the trailing libration island.
Both \cite{chiang2002} and \cite{murray-clay} found that the trailing 2:1 libration island should be more populated than the leading island, an effect that became more amplified with faster Neptune migration (up to $\sim300$\%).
However, the B\&M model has a factor of two enhancement in the leading island over the trailing island.
The preference for the leading libration island continues through all $n$:1 resonances that have enough test particles for a reliable ratio to be determined (though not all are statistically significant, see Table~\ref{n_outer}).
The B\&M simulation results confirm that migrations can produce asymmetries in libration island capture, however migration rate is not the only factor which can create an asymmetry.
The enhancement in particular libration islands must depend on additional migration parameters than purely migration speed, such as the starting orbital parameters of all of the giant planets and the initial orbits of the particles.

Measuring the libration island population fractions is a goal of characterized surveys; the complete OSSOS survey will provide strong constraints on this ratio \citep[e.g.][]{volk2016}.
Based on the first results from the OSSOS survey, \cite{volk2016} provides the only statistically robust constraint on the fraction of 2:1 asymmetric resonators in the leading island at $<90$\%; refining this estimate is a goal of the full OSSOS survey.
Determining relative population sizes of the different resonant components in the Kuiper belt based on survey results will provide constraints on different Solar System migration models.

More theoretical work is needed to understand the additional constraints which influence the specifics of where the asymmetries appear.
This is unfortunate as many authors have hoped that measuring the ratios of populations in the various libration islands could provide a straightforward diagnostic of migration speed; this result shows the interpretation will be more complicated.
Recent ``grainy'' migration simulations \citep{nesvorny2016,kaibsheppard2016} have shown promise in reproducing aspects of the observed Kuiper Belt, however, these works have not presented the resulting $n$:1 resonant island fractions so no direct comparisons can yet be made.

\subsection{Resonant Dropouts: The Importance of Kozai within MMRs}
\label{kozaistructure}

Recent simulations by \citet{nesvorny2016} and \citet{kaibsheppard2016} (hereafter referred to as N16 and KS16 respectively) present detailed analysis of resonance population by smooth and grainy Neptune migration models.  
Our Figure~\ref{allparticles} can be directly compared with Figure~2 in KS16 and Figures~1 and 2 in N16.
These Figures highlight the difference in the orbital distribution of near-resonant distant TNOs.

The main difference in near-resonant structure of the models is that the B\&M migration model produces a significant number of resonant dropouts at low $e$ on \emph{both} sides of each resonance.
Due to its relative isolation, this resonant `beard' is most easily visible by eye around the 3:1 resonance (Figure~\ref{allparticles}).
This relatively symmetric distribution of low-e test particles around each resonance happens because Neptune's very high eccentricity in the B\&M simulation initially produces wide, powerful resonances.  
As Neptune's orbit circularizes, and the resonances narrow, particles drop out of resonance and produce this `beard'-like structure \citep[as discussed in][]{gladman12}.  
In the smooth and grainy migration simulations in KS16 and N16, where Neptune stays below a more moderate $e$ of $\sim$0.1, the resulting features around resonances are not symmetric beard-like structures; there are more particles dropped on the sunward side of each resonance. 

\begin{figure}
\begin{center}
\includegraphics[width=0.5\textwidth]{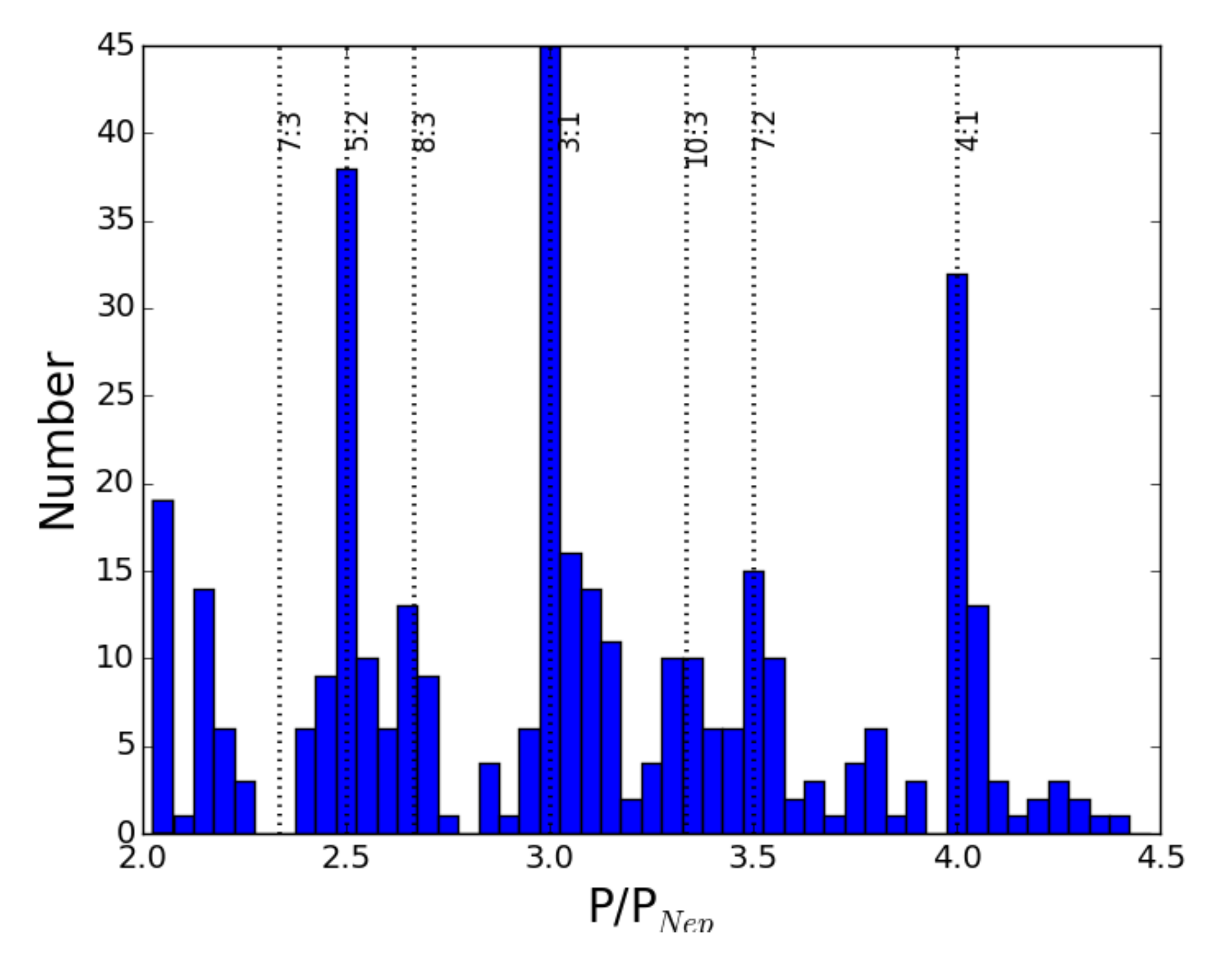}\includegraphics[width=0.5\textwidth]{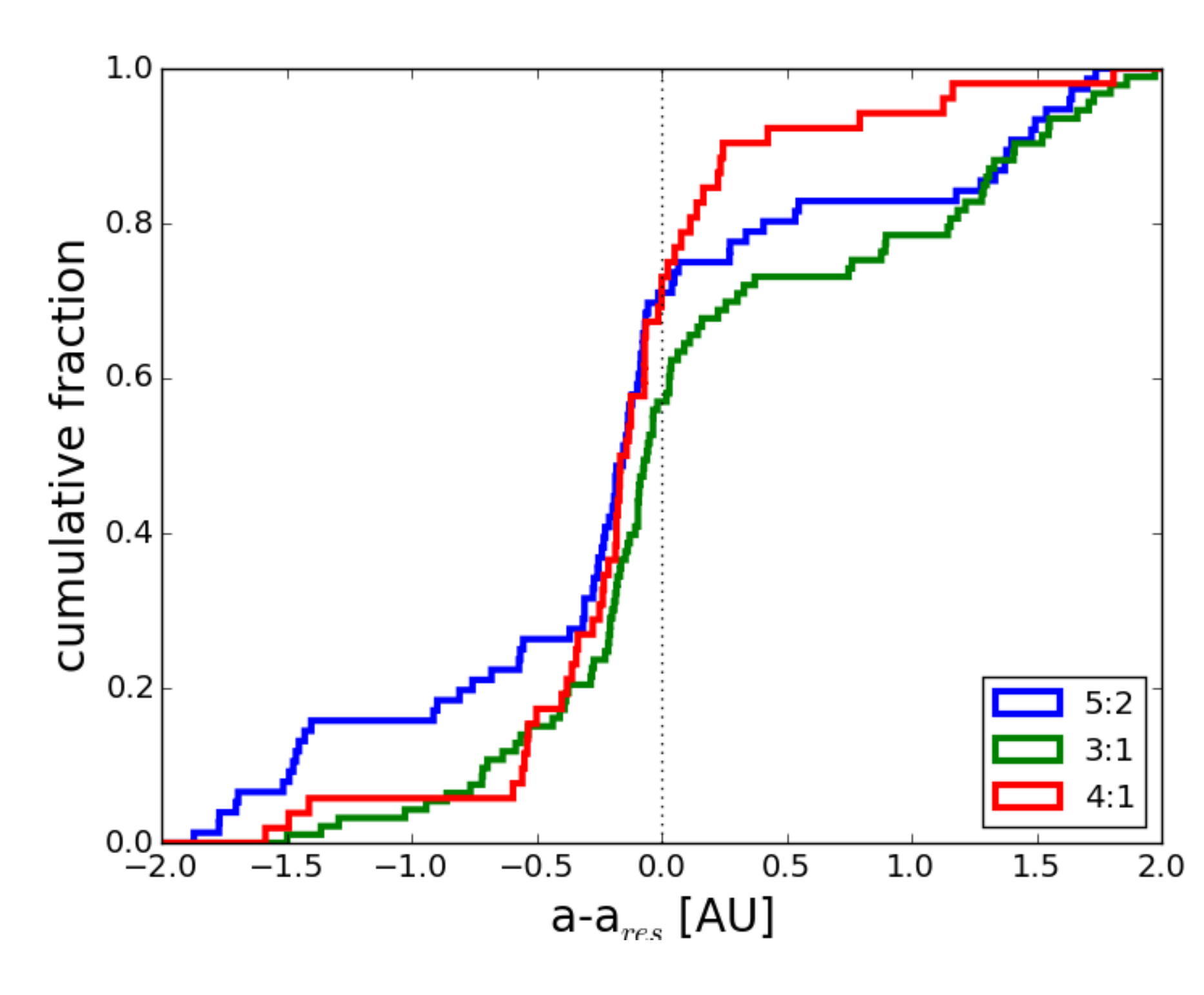}
\caption{
Details of high-$q$ ($q>40$~AU) test particles relative to Neptune MMRs.
\emph{Left:} Number of high-$q$ particles around several well-populated resonances \citep[compare to Figure~3 in][]{kaibsheppard2016}.  \emph{Right:} cumulative distributions of test particles around the 5:2, 3:1, and 4:1 MMRs \citep[compare to Figure~6 in][]{kaibsheppard2016}. 
The B\&M migration produces many particles Sunward of the resonances, but also has many on the outer side of the resonances.  
This is likely due to the higher $e$ attained by Neptune during this simulation as compared to the migration simulations in \citet{kaibsheppard2016}.
}
\label{natecompare}
\end{center}
\end{figure}

Figure~\ref{natecompare} highlights this near-resonant structure in the high-$q$ ($q>40$~AU) population.
(Compare this directly with Figures~3 and 6 in KS16.)
While all of the simulations presented in N16 and KS16 have many more particles located just Sunward of the resonances (due to resonant dropout during Neptune's migration phase), the B\&M simulation produces many particles on both sides of the resonances (Figure~\ref{natecompare}).  
This is likely due to a much more significant amount of dropout occurring during the circularization phase of Neptune's orbit evolution during the B\&M simulation, which attains much higher eccentricity during the Nice model instability ($e\sim0.3$) as compared with the N16 and KS16 simulations ($e\sim0.1$).

As discussed extensively in KS16, when exactly the low-$e$ dropouts primarily happen depends on the timescale of Neptune's orbital evolution relative to the Kozai cycling timescale within MMRs.  
In the N16 and KS16 analyses, more dropouts occur Sunward of the resonances in the slower migration simulations. 
This is because the timescale for particle capture and Kozai cycling is shorter than the slow migration timescale ($\sim$100~Myr) but longer than the fast migration timescale ($\sim$10~Myr).  

Particles are captured into Neptune's migrating resonances and then pushed to higher-$e$ by Neptune's continued outward migration, into phase space where Kozai oscillations are stronger.  
Kozai cycling will eventually drive the particles to higher $i$ and lower $e$, where the resonance is weaker, and dropout is more likely.  
How Neptune's orbit evolves on the timescale of a Kozai cycle within a MMR ($\sim$several Myr) will determine where low-$e$ dropouts are deposited.  
In the case of faster migration (10~Myr timescale simulations in N16 and KS16), Neptune's orbit migrates outward significantly before Kozai cycling can dramatically modify particle orbits, so the particles are likely to remain in resonance.
In the case of slower migration (100~Myr timescale N16 and KS16 simulations), Kozai cycling has time to modify particle orbits to lower-$e$ as the resonance slowly marches outward, resulting in preferential deposition of low-$e$ particles on the sunward side of resonances.
In the B\&M simulation, Neptune's orbital circularization is more significant than in the N16 and KS16 simulations, and occurs slowly enough that particles are Kozai cycled to lower-$e$ while the resonances narrow, leading to the more symmetric resonant beard structure we observe.

The low-$e$ Neptune simulations thus far do not contain nearly as many test particles as the B\&M simulation, so the details of resonance populations, in particular resonant island fractions, are not currently able to be compared in a statistically meaningful way.  
Since the B\&M simulation predicts a large leading-trailing asymmetry among the n:1 resonances, in the future it would be very interesting to compare this to predictions of resonant island occupation from these grainy/smooth lower $e$ simulations.

The resonant fingers in our Figure~\ref{allparticles} extending to low $e$ by eye most closely match the fluffier near-resonant structure visible in the N16 and KS16 simulations where Neptune had a slower migration timescale ($\sim$100~Myr).
It appears that the most important factor in determining how well-populated the near-resonant trails or beards are is the timescale of the evolution of Neptune's orbit relative to the Kozai cycling timescale for particles in Neptune's MMRs.
Future detailed observations of the populations of TNOs within and immediately surrounding these distant resonances will be an important and powerful diagnostic for the exact mode and timescale of Neptune's migration.

\section{Discussion and Conclusions}
\label{conclusions}

The B\&M Nice model simulation produces predictions for the distribution of TNOs in and surrounding distant mean motion resonances.
To summarize these predictions:
\begin{itemize}
\item Significant populations within $n$:1 and $n$:2 MMRs, even at very large (e.g. $>$120 AU) distances, with on-average higher $e$ and $i$ values attained in more distant resonances.
\item A lack of low-libration amplitude TNOs in $n$:2 resonances due to primarily scattering capture. 
\item More TNOs residing in the leading rather than the trailing asymmetric islands of all the $n$:1 resonances.
\item Roughly 20\% of TNOs in MMRs are also currently experiencing Kozai resonance.  
\item The high-$q$ ($q>40$~AU) population at distances beyond the 3:1 resonance should have overall higher inclinations than the low-$q$ population.  
\item The high-$q$ population immediately surrounding MMRs should show a fairly symmetric, beard-like distribution, having nearly as many TNOs just Sunward of the resonances as just beyond the resonances. This is in direct contrast to N16 and KS16 whose simulations predict a much larger population just Sunward of the resonances.
\end{itemize}
These predictions will be testable by the results of future well-characterized surveys containing hundreds of TNOs, such as OSSOS, which will provide evidence for or against the Nice model-style of Neptune migration.  

The B\&M simulation resulted in asymmetric capture into the $n$:1 resonance islands.
\cite{chiang2002} and \cite{murray-clay} determined that the {\bf trailing} libration island was more populated for the 2:1 resonators, captured through resonance sweeping.
The B\&M model produces an asymmetry in the 2:1 resonance, as well as the 3:1, 4:1, and 7:1 resonators (with less significant asymmetries in the 5:1 and 6:1 resonators), where the {\bf leading} libration island is more populated.
The particles trapped in the $n$:1 resonances in the B\&M simulations are captured through scattering; this shows that scattering capture with a large-$e$ Neptune also produces asymmetries.
The B\&M $n$:1 resonators display the opposite asymmetry than predicted by previous work, so asymmetries in libration island capture are indicative of specifics of migration, but capture asymmetry is sensitive to more than just the migration rate.
In addition to the capture mechanism, the initial planet and test particle semi-major axes and eccentricities may also play a role, and more work is needed to determine the best interpretation of asymmetries in the $n$:1 populations if they are discovered in future observations of the real Kuiper belt.

The effect of the Kozai mechanism in the simulations is apparent in multiple populations.
Approximately 20\% of the 3:2, 2:1, and 5:2 resonators are in Kozai at the end of the simulation, which matches the measured Kozai fraction measured for the Plutinos \citep{volk2016}.
\cite{Lykawka_Mukai2007Kozai} found a Kozai fraction 20-30\% for the 2:1's, consistent with the B\&M fraction, but observation biases for Kozai resonators are complicated, and a complete model is required to properly debias these observed Kozai fractions \citep{lawler13}.
The signature of the Kozai effect is also identifiable in non-resonant populations; these particles were likely resonant and in Kozai at an earlier time in the B\&M orbital evolutions.
Particles can enter Kozai more easily while in mean motion resonance since Kozai oscillations occur at much lower inclinations inside resonances. 
This alternately pumps the particle's eccentricity or inclination.
At large-$e$ and large-$i$, resonances are less strong, and particles can diffuse or drop out of resonance.

The non-resonant simulated particles beyond the 3:1 resonance have distinct $e$ and $i$ distributions.
The high-$q$ ($q>40$~AU) objects have lower $e$ (resulting in the large $q$), but this low $e$ is strongly correlated with a hotter inclination distribution.
These large inclinations and low eccentricities were likely produced through temporary capture into a mean-motion resonance and Kozai, which decreased the eccentricity and increased the inclination.
The test particles then diffused or dropped out of the resonance \citep{Gomes2011} and became detached or outer classical objects.
The detached objects include a large fraction of these high-$q$ objects, and this population had almost as cold an $e$ and hot an $i$ distribution, but is diluted by objects which are not significantly evolved by Kozai resonance capture.
The scattering objects are those that had the opposite occur: they are likely created by leaking out of resonance after the $e$ is increased and the $i$ is decreased due to Kozai inside mean-motion resonances.

The resonant fingers descending into low-$e$, high-$q$ at each resonance are the result of Kozai cycling.
The timescale of Neptune's migration and circularization will affect whether the low-$q$ population just outside the resonances will be primarily Sunward (as in the N16 and KS16 slower migrations), will be very sparse (as in the N16 and KS16 faster migrations), or well-populated and distributed fairly evenly on both sides of the resonances (as in this simulation).

Upcoming large surveys will detect more high-$q$ TNOs.
However, only carefully characterized surveys where the biases are well-known and published, will be useful for measuring populations and structures within these distant resonances, where observing biases are extreme and complicated \citep[e.g.][]{volk2016}.
OSSOS is just such a well-characterized survey, and upcoming analysis will provide important constraints on Neptune's migration based on these and other predictions in the literature.

\acknowledgements
The authors thank R.~Brasser for providing the output of his simulation for this analysis, C.~Shankman for providing his resonance diagnostic code, N.~Kaib for sharing his migration code and simulations, J.J.~Kavelaars for helpful discussions, and an anonymous referee for comments which helped improve this manuscript.
SML gratefully acknowledges support from the NRC-Canada Plaskett Fellowship.  This research used the facilities of the Canadian Astronomy Data Centre operated by the National Research Council of Canada with the support of the Canadian Space Agency.  The authors wish to acknowledge the important (though distracting) role played by the babies they each had during the preparation of this manuscript, and thank their respective institutions for providing paid parental leave.

\software{Python, Swift}


\begin{thebibliography}{}
\expandafter\ifx\csname natexlab\endcsname\relax\def\natexlab#1{#1}\fi

\bibitem[{{Alexandersen} {et~al.}(2016){Alexandersen}, {Gladman}, {Kavelaars},
  {Petit}, {Gwyn}, {Shankman}, \& {Pike}}]{alexandersen}
{Alexandersen}, M., {Gladman}, B., {Kavelaars}, J.~J., {et~al.} 2016, \aj

\bibitem[{Anderson \& Darling(1954)}]{andersondarling54}
Anderson, T.~W., \& Darling, D.~A. 1954, Journal of the American Statistical
  Association, 49, 765
  
\bibitem[Bannister et al.(2016)]{Bannisteretal2016} 
Bannister, M.~T., Kavelaars, J.~J., Petit, J.-M., et al.\ 2016, \aj, 152, 70 

\bibitem[{{Batygin} {et~al.}(2011){Batygin}, {Brown}, \&
  {Fraser}}]{batygin2011}
{Batygin}, K., {Brown}, M.~E., \& {Fraser}, W.~C. 2011, \apj, 738, 13

\bibitem[{{Brasser} \& {Morbidelli}(2013)}]{brasser2013}
{Brasser}, R., \& {Morbidelli}, A. 2013, \icarus, 225, 40

\bibitem[{{Brown}(2001)}]{brown2001}
{Brown}, M.~E. 2001, \aj, 121, 2804

\bibitem[{{Chiang} \& {Jordan}(2002)}]{chiang2002}
{Chiang}, E.~I., \& {Jordan}, A.~B. 2002, \aj, 124, 3430

\bibitem[{{Gladman} {et~al.}(2008){Gladman}, {Marsden}, \&
  {Vanlaerhoven}}]{gladman2008}
{Gladman}, B., {Marsden}, B.~G., \& {Vanlaerhoven}, C. 2008, {Nomenclature in
  the Outer Solar System} (The University of Arizona Press), 43--57

\bibitem[{{Gladman} {et~al.}(2012){Gladman}, {Lawler}, {Petit}, {Kavelaars},
  {Jones}, {Parker}, {Van Laerhoven}, {Nicholson}, {Rousselot}, {Bieryla}, \&
  {Ashby}}]{gladman12}
{Gladman}, B., {Lawler}, S.~M., {Petit}, J.-M., {et~al.} 2012, Astronomical
  Journal, 144, 23

\bibitem[{{Gomes}(2003)}]{gomes2003}
{Gomes}, R.~S. 2003, \icarus, 161, 404

\bibitem[Gomes et al.(2005)]{Gomes2005} 
Gomes, R.~S., Gallardo, T., Fern{\'a}ndez, J.~A., \& Brunini, A.\ 2005, Celestial Mechanics and Dynamical Astronomy, 91, 109 


\bibitem[{{Gomes} {et~al.}(2008){Gomes}, {Fern Ndez}, {Gallardo}, \&
  {Brunini}}]{gomes_ssbn}
{Gomes}, R.~S., {Fern Ndez}, J.~A., {Gallardo}, T., \& {Brunini}, A. 2008, {The
  Scattered Disk: Origins, Dynamics, and End States}, ed. M.~A. {Barucci},
  H.~{Boehnhardt}, D.~P. {Cruikshank}, A.~{Morbidelli}, \& R.~{Dotson},
  259--273
  
\bibitem[Gomes(2011)]{Gomes2011} 
Gomes, R.~S.\ 2011, \icarus, 215, 661 

\bibitem[{{Gulbis} {et~al.}(2010){Gulbis}, {Elliot}, {Adams}, {Benecchi},
  {Buie}, {Trilling}, \& {Wasserman}}]{gulbis}
{Gulbis}, A.~A.~S., {Elliot}, J.~L., {Adams}, E.~R., {et~al.} 2010, \aj, 140,
  350
  
\bibitem[Kaib \& Sheppard(2016)]{kaibsheppard2016} 
Kaib, N.~A., \& Sheppard, S.~S.\ 2016, \aj, 152, 133

\bibitem[{{Kozai}(1962)}]{kozai2}
{Kozai}, Y. 1962, \aj, 67, 591

\bibitem[{{Lawler} \& {Gladman}(2013)}]{lawler13}
{Lawler}, S.~M., \& {Gladman}, B. 2013, \aj, 146, 6

\bibitem[{{Levison} {et~al.}(2008){Levison}, {Morbidelli}, {Van Laerhoven},
  {Gomes}, \& {Tsiganis}}]{nice}
{Levison}, H.~F., {Morbidelli}, A., {Van Laerhoven}, C., {Gomes}, R., \&
  {Tsiganis}, K. 2008, \icarus, 196, 258

\bibitem[{{Lidov}(1962)}]{kozai1}
{Lidov}, M.~L. 1962, \planss, 9, 719

\bibitem[{{Lykawka} \& {Mukai}(2007)}]{Lykawka_Mukai2007Kozai}
{Lykawka}, P.~S., \& {Mukai}, T. 2007, \icarus, 189, 213

\bibitem[Lykawka \& Mukai(2007)]{LykawkaMukai2007Scattering} 
Lykawka, P.~S., \& Mukai, T.\ 2007, \icarus, 192, 238 

\bibitem[Malhotra(1993)]{Malhotra1993} 
Malhotra, R.\ 1993, \nat, 365, 819 

\bibitem[Morbidelli et al.(2005)]{morbidelli2005} Morbidelli, A., Levison, H.~F., Tsiganis, K., \& Gomes, R.\ 2005, \nat, 435, 462 


\bibitem[{{Murray-Clay} \& {Chiang}(2005)}]{murray-clay}
{Murray-Clay}, R.~A., \& {Chiang}, E.~I. 2005, \apj, 619, 623

\bibitem[{{Nesvorn{\'y}}(2015{\natexlab{a}})}]{nesvorny2015b}
{Nesvorn{\'y}}, D. 2015{\natexlab{a}}, \aj, 150, 73

\bibitem[{{Nesvorn{\'y}}(2015{\natexlab{b}})}]{nesvorny2015a}
---. 2015{\natexlab{b}}, \aj, 150, 68

\bibitem[Nesvorn{\'y} et al.(2016)]{nesvorny2016} 
Nesvorn{\'y}, D., Vokrouhlick{\'y}, D., \& Roig, F.\ 2016, \apjl, 827, L35

\bibitem[{{Petit} {et~al.}(2011){Petit}, {Kavelaars}, {Gladman}, {Jones},
  {Parker}, {Van Laerhoven}, {Nicholson}, {Mars}, {Rousselot}, {Mousis},
  {Marsden}, {Bieryla}, {Taylor}, {Ashby}, {Benavidez}, {Campo Bagatin}, \&
  {Bernabeu}}]{cfeps}
{Petit}, J.-M., {Kavelaars}, J.~J., {Gladman}, B.~J., {et~al.} 2011, \aj, 142,
  131

\bibitem[{{Pike} {et~al.}(2015){Pike}, {Kavelaars}, {Petit}, {Gladman},
  {Alexandersen}, {Volk}, \& {Shankman}}]{pike2015}
{Pike}, R.~E., {Kavelaars}, J.~J., {Petit}, J.~M., {et~al.} 2015, \aj, 149, 202

\bibitem[Pike et al.(2017)]{pike2017} Pike, R.~E., Lawler, S., Brasser, R., et al.\ 2017, \aj, 153, 127 

\bibitem[Shankman et al.(2016)]{Shankman2016} Shankman, C., Kavelaars, J., Gladman, B.~J., et al.\ 2016, \aj, 151, 31

\bibitem[{{Shankman} {et~al.}(In Prep){Shankman}, {Am{\'e}lard}, {Kavelaars},
  {Pike}, \& {Wong}}]{shankmanFFT}
{Shankman}, C., {Am{\'e}lard}, R., {Kavelaars}, J., {Pike}, R.~E., \& {Wong},
  A., In prep.

\bibitem[{{Thomas} \& {Morbidelli}(1996)}]{thomas1996}
{Thomas}, F., \& {Morbidelli}, A. 1996, Celestial Mechanics and Dynamical
  Astronomy, 64, 209

\bibitem[Tiscareno \& Malhotra(2009)]{TiscarenoMalhotra2009} 
Tiscareno, M.~S., \& Malhotra, R.\ 2009, \aj, 138, 827 

\bibitem[Tsiganis et al.(2005)]{NiceModel2005} 
Tsiganis, K., Gomes, R., Morbidelli, A., \& Levison, H.~F.\ 2005, \nat, 435, 459

\bibitem[{{Volk} {et~al.}(2016){Volk}, {Murray-Clay}, {Gladman}, {Lawler},
  {Bannister}, {Kavelaars}, {Petit}, {Gwyn}, {Alexandersen}, {Chen}, {Lykawka},
  {Ip}, \& {Lin}}]{volk2016}
{Volk}, K., {Murray-Clay}, R., {Gladman}, B., {et~al.} 2016, \aj, 152, 23

\bibitem[{{Williams} \& {Benson}(1971)}]{williams1971}
{Williams}, J.~G., \& {Benson}, G.~S. 1971, \aj, 76, 167

\end{thebibliography}
\end{document}